\documentclass[seceq]{ptptex}

\usepackage{graphicx}

\def\ignore#1{{}}

\newcommand{\beeq}{\begin{equation}}
\newcommand{\eneq}{\end{equation}}
\newcommand{\beqn}{\begin{eqnarray}}
\newcommand{\eeqn}{\end{eqnarray}}

\def\la{\raise.16ex\hbox{$\langle$}\lower.16ex\hbox{}  }
\def\ra{\, \raise.16ex\hbox{$\rangle$}\lower.16ex\hbox{} }
\def\go{\rightarrow}

\def\next{{~~~,~~~}}
\def\onehalf{ \hbox{${1\over 2}$} }

\def\Pl{{\rm Pl}}

\def\ep{\epsilon}
\def\psibar{ \psi \kern-.65em\raise.6em\hbox{$-$} \lower.6em\hbox{} }
\def\psibarb{ \psi \kern-.65em\raise.6em\hbox{$-$}  }
\def\Rbar{{\overline R}}

\title{%        %You can use \\ for explicit line-break
Gravitating Fermionic Lumps with a False Vacuum Core
}

\author{%       %Use \scshape  for the family name
Ramin G.\ \textsc{Daghigh}$^{1}$ and
 Yutaka \textsc{Hosotani}$^{2}$}%

\inst{%         %Affiliation, neglected when [addenda] or [errata]
$^1${Department of Applied Mathematics \& Theoretical Physics,
The Queen's University of Belfast, Belfast, BT7 1NN, Northern Ireland, UK}
\\
$^2${ Department of Physics, Osaka University,
Toyonaka, Osaka 560-0043, Japan}
}

\abst{
We investigate gravitating lumps with a false vacuum core surrounded 
by the true vacuum in a scalar field potential. Such configurations become
possible in  Einstein gravity in the presence of
fermions at the core.  Gravitational interactions as well as Yukawa
interactions are essential for  such  lumps to exist.  
The mass and size of gravitating lumps sensitively depend on the scale
characterizing  the scalar field potential  and the density of fermions.
These objects can exist in the universe at various scales.  
}

\begin{document}

\maketitle

\section{Introduction}

When a scalar field potential has two non-degenerate minima,  the absolute
minimum of the potential corresponds to the true vacuum, while the other 
to the false vacuum.  Systems in which false and true vacua 
coexist are of great interest.
The universe is full of various structures such as
black holes, stars and dark matter. Does there exist 
any structure that has a false vacuum in its core? 
If the entire universe is in a false vacuum, it decays into
the true vacuum through bubble creation by quantum tunneling.\cite{Coleman1}
What happens if a false vacuum core is surrounded by the true 
vacuum?\cite{Daghigh, Hosotani2}
If the size of the core is smaller than the critical radius, the core
would quickly decay, with its energy dissipating to the spatial infinity.
If the size of the core is larger than the critical radius, the 
core becomes a black hole.  In either case the configuration cannot
be static.  The fate of false-vacuum bubbles has been extensively discussed
in the literature.\cite{Blau, NoGo, Galtsov}

It has been shown recently that  new configurations, cosmic shells, emerge in
a simple real scalar field theory in which spherical shells of the true vacuum
are immersed in the false vacuum.\cite{Hosotani1}  Such static cosmic shells
can  exist thanks to gravitational interactions.  However, a static ball of the
false vacuum immersed in the true vacuum is not possible.

In this paper we  demonstrate that 
a static false vacuum core becomes possible if there is
additional matter coupled to the scalar field such as fermions.
The number of fermions necessary is not large.  
Gravitational interactions as well as Yukawa interactions play a key role
in making such a structure possible. 
It is a gravitating fermionic lump. We stress that  gravitating lumps
are  quite different from Q-balls, boson stars and Fermi-balls. In Q-balls   
the conserved  charge of the scalar field plays a  central role in realizing
the stability.\cite{Qball}  There is no such charge of the scalar field
in our model.  In boson stars,  gravitational interactions as well as the 
conserved charge play a key role.\cite{BosonStar1, BosonStar2}
The model for Fermi balls is very similar to
ours \cite{Campbell, Arafune} in which the false vacuum is surrounded by the
true vacuum.  In Fermi balls,  fermions are localized in the transition
region, or the domain wall, i.e.\ they reside in the surface region of the
lumps.  In our model fermions reside in the bulk region inside the lump and
the Yukawa interaction becomes essential.

Gravitational interactions can produce lump
solutions in non-Abelian gauge theory as well.  Even in the pure 
Einstein-Yang-Mills theory stable monopole solutions appear in 
the asymptotically anti-de Sitter space.\cite{Bjoraker, Bartnik, Torii}
We  show in the present paper that such gravitational lumps appear
even in a simple scalar field theory with fermions.

The spacetime geometry of gravitating lumps is either anti-de
Sitter-Schwarzschild or de Sitter-Schwarzschild.   Dymnikova has discussed 
the global structure of de Sitter-Schwarzschild  spacetime, assuming an
appropriate external matter distribution.\cite{Dymnikova}  The existence of
gravitating lumps  in the present paper shows that  anti-de
Sitter-Schwarzschild spacetime is indeed realized in a very simple system. 
Its existence can be inferred from the energetics consideration in flat 
spacetime as well.  The extension to  de Sitter-Schwarzschild 
 spacetime is reserved for future investigation.   One may explore such 
objects in the universe at various scales and epochs.  We shall see
that the size of gravitating lumps sensitively depends on the scale
characterizing the scalar field potential.

We stress that the gravitating fermionic lumps  described in the present
paper exist when there appears a false vacuum and there are fermions
coupled to the relevant scalar field.  Recently it has been shown that such
a false vacuum  appears in the early universe in the standard 
Einstein-Weinberg-Salam theory of electroweak and gravitational 
interactions, if the universe has a spatial section $S^3$ as in the 
closed Friedmann-Robertson-Walker universe.\cite{Emoto}   In the early
stage, nontrivial gauge fields yield a false vacuum in the Higgs field,
to which quarks and leptons have  Yukawa couplings.  In other 
words gravitating fermionic lumps may be copiously produced in the
framework of the standard  model.

In \S 2 we set up the problem and introduce useful
variables in terms of which the field equations are written. 
In \S 3 we give the energetics of fermionic lumps, based mainly 
on flat spacetime.  This illustrates how such lumps become possible
when the scalar field potential has a false vacuum configuration.
In \S 4 the behavior of the solutions is investigated 
analytically inside the lump, numerically in the transition region,
and analytically outside the lump.  More details of the solutions
are given in \S 5 with a focus on the dependence of
the  solutions on various parameters of the model.   
It is seen how the solutions change as 
the energy scale of the model is lowered. 
A summary and conclusions are given in \S 6.

\section{Model}

\begin{figure}[tb]
\begin{center}
\includegraphics[height=5cm]{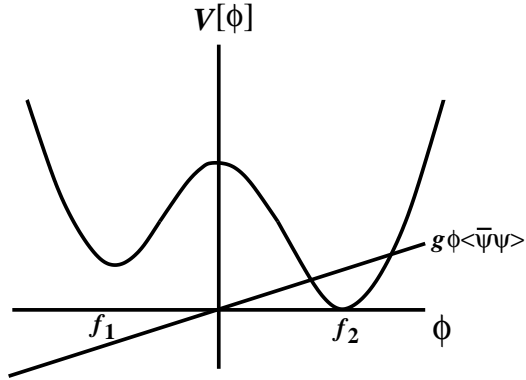}
% %potential.eps = fig02.eps
\end{center}
\caption{A scalar field potential $V[\phi]$ with two minima.  The quantity
$\phi \rho_S=g\phi\la \psibarb \psi\ra$  at the core is also displayed
for $g \la \psibarb \psi\ra > 0$.   In the lump solutions described 
in the present paper, $g |f_1| \la \psibarb \psi\ra$ is much larger than 
the energy density of the false vacuum at $f_1$.}
\label{potential}
\end{figure}

We consider a real scalar field in the Einstein gravity whose Lagrangian 
is given by 
\beeq
\mathcal{L}=\frac{1}{16\pi G}R+\frac{1}{2}g^{\mu\nu}\partial_\mu\phi
\partial_\nu\phi-V[\phi]- \phi \rho_S(x)
\label{model1}
\eneq
where \(R\) and  \(V[\phi]\)  are the scalar curvature and the scalar 
potential, respectively.   The last term in (\ref{model1}) 
represents a source for the scalar field. It naturally arises
if there is a fermion field $\psi$ with a Yukawa interaction
$- g \phi \bar\psi \psi$.\cite{Hashimoto}  In this case 
$\rho_S(x) = g \la \psibar \psi \ra$.  
\ignore{$\psi$ can be either a Dirac or Majorana field. (??)}
As shown schematically
in Fig.\ \ref{potential},  \(V[\phi]\) has two minima at $f_1$ and \(f_2\).
We take
\begin{eqnarray}
V[\phi] \, &=& \frac{\lambda}{4} (\phi-f_2)
\Bigg\{ \phi^3-\frac{1}{3}(f_2+4f_1)\phi^2
  - \frac{1}{3}f_2(f_2-2f_1)(\phi+f_2) \Bigg\}  ~, \cr
\noalign{\kern 10pt}
V'[\phi]&=&\lambda\phi(\phi-f_1)(\phi-f_2)  ~.
\label{potential1}
\end{eqnarray}

In the potential \(V[\phi]\), the false vacuum appears at $\phi=f_1$
and the true vacuum at  $\phi =f_2$.  We investigate spherically
symmetric configurations in which the scalar field varies from the
false vacuum  to the true vacuum as the radius $r$ increases.
Solutions $\phi(r)$ are to satisfy
the two conditions ({\it i}) $\phi(0) \simeq f_1$ and ({\it ii})
$\phi(\infty) = f_2$.

In (\ref{model1}) we have added the interaction term 
$- \phi \rho_S(x)$ to the scalar field. This interaction 
is  vital to have solutions satisfying  the condition (ii). 
As we shall see below, solutions regular at the origin are uniquely
specified by the value of $\phi(0)$ with a given $\rho_S(x)$.  
Without the coupling to a source
$- \phi \rho_S(x)$, the configuration $\phi(r)$  either settles at $\phi=0$
after oscillation as $r \go \infty$, which corresponds to a totally unstable
configuration, or comes back to $\phi= f_1$, or overshoots $\phi= f_1$ to
diverge to $-\infty$ as $r$ approaches $\infty$.  
Solutions of the second type indeed exist.  They are  cosmic
shells.\cite{Hosotani1}
The source term $- \phi \rho_S(x)$ gives a $x$-dependent linear
potential for the scalar field. (See Fig.\ \ref{potential}.)
\ignore{
Although it is vital, its magnitude
can be relatively small.  Even with this interaction the energy density 
of the false vacuum remains higher than that of the true vacuum. 
The gravitational interaction amplifies the effect of the force
applied by the source. (?)
}

We treat the system classically.   If the source comprises fermions,
$\rho_S(x) = g \la \psibar \psi(x) \ra$ where $g$ is a Yukawa coupling
constant.  If the fermions at the core are nonrelativistic,   
$\la \psibar \psi(x) \ra$ can be approximated by  the fermion density  
$\la \psi^\dagger \psi(x) \ra \equiv \rho_n(x)$.  In the present paper
we suppose that the fermion density is small enough so that the 
back-reaction to the metric can be ignored.  This condition is met in a
simple model if
the fermion couples to a pair of scalar fields in the specific manner 
described in the next section.   A full treatment of  the dynamics of the
sources including back-reaction to the Einstein equations is left for
a future study.

We look for spherically symmetric, static 
configurations. The metric of spacetime is written as 
\begin{equation}
ds^2=\frac{H}{p^2}dt^2-\frac{dr^2}{H}-r^2(d\theta^2+\sin^2\theta 
d\varphi^2) 
\label{eq:1a}
\end{equation}
where  \(H\) and \(p\) are  functions depending only on \(r\).
In the tetrad basis
\beeq
e_0=\frac{\sqrt{H}}{p}dt ~,~~ e_1=\frac{1}{\sqrt{H}}dr ~,~~
e_2=rd\theta ~,~~  e_3=r\sin\theta d\varphi ~, 
\eneq
the energy-momentum tensors $T_{ab}=e_a^{\ \mu} e_b^{\ \nu} T_{\mu\nu}$
are 
\beqn
&&\hskip -1cm 
T_{00} =\frac{1}{2}H{\phi'}^2
+V[\phi]+ \phi \rho_S ~, \cr
\noalign{\kern 8pt}
&&\hskip -1cm 
T_{11} = \frac{1}{2}H{\phi'}^2
-V[\phi]-  \phi \rho_S ~, \cr
\noalign{\kern 8pt}
&&\hskip -1cm 
T_{22} = T_{33}= -\frac{1}{2}H{\phi'}^2
        -V[\phi]-  \phi \rho_S ~, \cr
\noalign{\kern 8pt}
&&\hskip -1cm 
\hbox{others}  = 0 ~~, 
\label{em-tensors}
\end{eqnarray}
where  the prime represents a  \(r\)-derivative. 

The Einstein equations are 
\begin{eqnarray}
-\frac{H'}{r}+\frac{1-H}{r^2}&=&8\pi G \, T_{00} ~, 
\label{Ein1}  \\
\frac{1}{r}\Bigl(-\frac{2Hp'}{p}+H'\Bigr)-\frac{1-H}{r^2}
  &=&8\pi G \, T_{11} ~, 
\label{Ein2}  \\
\frac{p}{2}\frac{\partial}{\partial
r}\Bigl(\frac{H'}{p}-\frac{2Hp'}{p^2}\Bigr)
-\frac{Hp'}{rp}+\frac{H'}{r}&=&8\pi G \, T_{22} ~.  
\label{Ein3}
\end{eqnarray}
It follows from (\ref{Ein1}) that 
\begin{equation}
H=1-\frac{2GM}{r} ~~,~~
M=\int_0^r 4\pi r^2  dr \, T_{00} ~~.
\label{metric2}
\end{equation}
Adding (\ref{Ein1}) to (\ref{Ein2}), we find
\begin{equation}
\frac{p'}{p}=-4\pi Gr{\phi'}^2 ~~.
\label{metric3}
\end{equation}
$p(r)$ is a monotonically decreasing function.
The equation of motion for the scalar field is 
\begin{equation}
-\frac{p}{r^2}\frac{\partial}{\partial r}
\Bigl(\frac{r^2H}{p}\phi'\Bigr)+V'[\phi]+ \rho_S=0
\label{scalar1}.
\end{equation} 
Eq.\ (\ref{Ein3}) follows from Eqs.\ (\ref{Ein1}), (\ref{Ein2}),
and (\ref{scalar1}).  

After eliminating $p(r)$, the equations are reduced
to (\ref{metric2}) and 
\begin{equation}
\phi''(r)+\bigg( \frac{2}{r}+4\pi Gr 
   {\phi'(r)}^2+\frac{H'}{H} \bigg) \phi'(r)
=\frac{1}{H}\Big (V'[\phi]+ \rho_S  \Big)  ~~.
\label{scalar2} 
\end{equation}
In interpreting Eq.\ (\ref{scalar2}) it is convenient to
introduce the new coordinate $\xi$ as
\beeq
\xi (r) = \int_0^r {dr \over \sqrt{ H }} ~~.
\label{new-xi}
\eneq
In the examples discussed below $H(r)$ remains positive definite so that
$\xi(r)$ is well defined.  Then the equation reads
\beqn
&&\hskip -1cm
{d^2 \phi\over d\xi^2} + \Gamma(\xi) {d \phi\over d\xi}
 = V'[\phi]+ \rho_S(\xi) ~~,
\label{scalar3} \\
&&\hskip -1cm
\Gamma(\xi) = \sqrt{H} \bigg( \frac{2}{r}+4\pi Gr 
           {\phi'(r)}^2+\frac{H'}{2H} \bigg)  ~~.
\eeqn
In terms of the ``time'' coordinate $\xi$ a particle with position
$\phi(\xi)$ moves in the potential $-V[\phi]$ with a time-dependent
external force $\rho_S(\xi)$.  There is friction $\Gamma$ which can be
either positive or negative.  The solution we are looking for corresponds 
to a particle which starts from $\sim f_1$ and moves to $f_2$
asymptotically.  It must gain an energy as $-V[f_1] < -V[f_2]$, which
is possible as there is an additional force given   by $\rho_S$ and 
$\Gamma$ can be negative.

\begin{figure}[htb]
\begin{center}
\includegraphics[height=5cm]{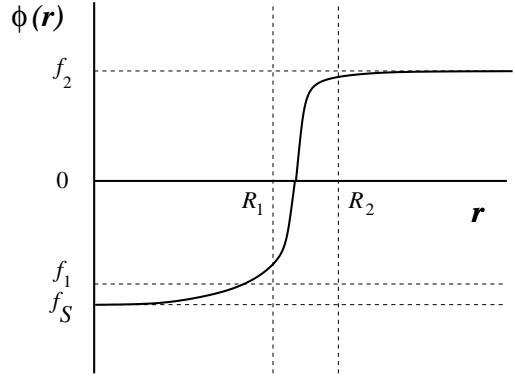}
\end{center}
\caption{The behavior of $\phi(r)$.  In the lump solutions 
$|f_1 - f_S| \ll |f_1|$ and $R_2-R_1 \ll R_1$.}
\label{phi-1}
\end{figure}

Equations (\ref{metric2})  and (\ref{scalar2}) define a set of  nonlinear
equations.   To be definite we suppose that the source to $\phi$
is given by
\beeq
\rho_S(r) = \rho_0 \theta(R_1 - r) ~~.
\label{f-radius}
\eneq
We split the space into three regions:
\beqn
{\rm I.} \quad && 0 \le r \le R_1 \cr
{\rm II.} \quad && R_1 \le r \le R_2  \cr
{\rm III.} \quad && R_2 \le r < \infty
\nonumber
\eeqn
\noindent
In region I, the spacetime is approximately anti-de Sitter with $\rho_S >0$ 
and $\phi_0$ very close to, but still greater than, the location $f_S$ of
the minimum of $V[\phi] + \phi \rho_S$: $\phi(0) = f_S + \delta \phi(0)$
with 
$0 < \delta\phi(0)/|f_S| \ll 1$.  It turns
out that
$\phi(r)$ varies little from 
$f_S$ in region I so that the equation of motion for $\phi$ may be 
linearized in this region.  In region II, $\rho_S=0$.  In this region the
field varies  significantly so that  the  full set of nonlinear equations
must be solved numerically.  In region III, $\rho_S=0$ and the spacetime is
approximately Schwarzschild.  In this paper we focus on the case in which
$|f_1|\sim f_2 \sim f \equiv \onehalf (|f_1| + f_2)$, 
$f_r = (f_2 - |f_1|)/f \ll 1$  and 
$|f_1-f_S| \ll f$  so that the linearization of   Eq.\ (\ref{scalar2})
is valid.
The behavior of  a solution $\phi(r)$ is displayed schematically in 
Fig.\  \ref{phi-1}.

\begin{figure}[htb]
\begin{center}
\includegraphics[height=5cm]{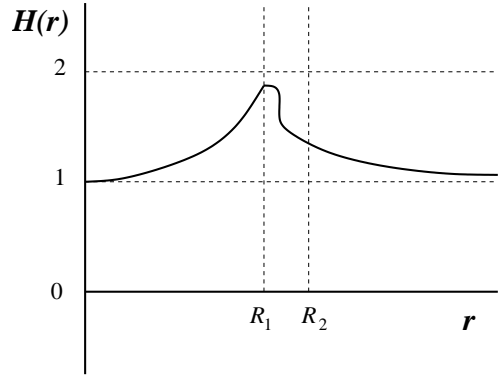}
\end{center}
\caption{The behavior of $H(r)$.  $H'(r)$ is discontinuous at $R_1$
as $T_{00}$ is discontinuous there when $\rho_S$ is given by
(\ref{f-radius}).}
\label{H-1}
\end{figure}

\begin{figure}[htb]
\begin{center}
\includegraphics[height=5cm]{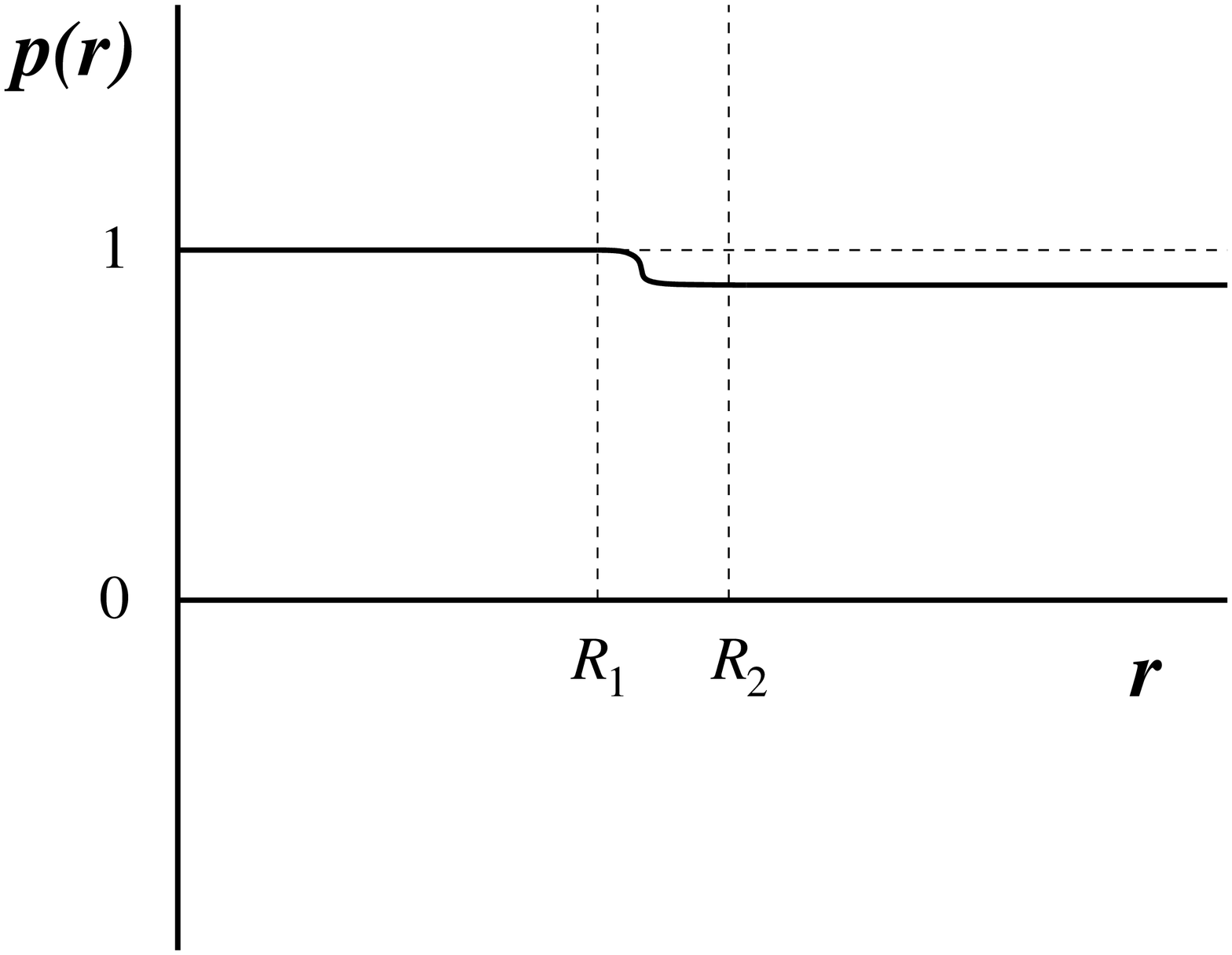}
\end{center}
\caption{The behavior of $p(r)$.}
\label{p-1}
\end{figure}

%\vskip 2cm

\section{Energetics of fermionic lumps}

It is instructive to see how fermionic lumps become possible
from  consideration of  energetics in flat spacetime.  The basic idea is
that there reside nonrelativistic fermions at the core, which generate
an additional linear interaction $\phi \rho_S$.  For nonrelativistic 
fermions $\la \psibar \psi \ra \sim \la \psi^\dagger \psi \ra 
\equiv \rho_n$ so that the Yukawa interaction $g \phi \psibar \psi$
generates a linear potential $g \rho_n \phi$ for the scalar field
inside the lump.  Let $R$ be the radius of the lump.  We suppose that
the total fermion number $N_F$ is conserved.
$N_F$ is kept fixed in the consideration below.

The scalar potential $V[\phi]$ takes the form depicted in Fig.\
\ref{potential}. The total potential $V[\phi] + g \rho_n \phi$ has a
minimum at $f_S <0$. It is supposed that $V[f_1] \equiv \ep_0 > 0$ but 
the energy density at the minimum  $V[f_S] + g \rho_n f_S =  \ep <0$ 
becomes negative for the lump solution.
$f_S$ and $\ep$ depends on $\rho_n$ or equivalently on $R$.

Inside the lump $r < R$, the scalar field satisfies $\phi \sim f_S$ whereas
outside the lump $\phi = f_2$.  A degenerate nonrelativistic fermion
gas has energy density ${\cal E}_f = mc^2 \rho_n + A \rho_n^{5/3}$ where
$A= 3 (3\pi^2)^{2/3} \hbar^2 / 10 m$.  
The total energy of the lump is
approximately given by 
\beeq
E(R) = \Big\{ {\cal E}_f + \ep (\rho_n) \Big\}  \, {4\pi R^3\over 3} 
 + 4\pi R^2 \sigma + E_{f, \rm wall}(R)
\label{energy1}
\eneq
where $\sigma$ is the surface tension resulting from varying $\phi$ in
the boundary wall region.  $E_{f, \rm wall}(R)$ is the contribution to the
energy from fermions localized in the  boundary wall region. It has been
estimated in ref.\ \cite{Arafune} to be about 
$2N_{f, \rm wall}^{3/2} / 3R$
where $N_{f, \rm wall}$ is the number of fermions confined in the boundary
wall region. 

Let the total fermion number $N_F \sim (4\pi R^3/3) \rho_n + N_{f, \rm wall}$ 
be fixed.    $\ep (\rho_n) \sim \ep_0$ for $\rho_n \sim 0$ or as $R \go \infty$.
For large $R$,  $E(R) \sim (4\pi \ep_0 /3) R^3$. 
For large $\rho_n$ (small $R$) the total potential is approximated by
$(\lambda/4) \phi^4 + g \rho_n \phi$ near the minimum so that
$\ep  \sim - {3\over 4} \lambda^{-1/3} (g \rho_n)^{4/3}$.  
When $N_F \sim (4\pi R^3/3) \rho_n$, 
 the ${\cal E}_f$-term dominates over the $\ep$-term and
$E(R) \sim A' R^{-2}$ ($A' >0$) for small $R$.  
In the case that $\rho_n$ remains small for small $R$,
$E_{f, \rm wall}(R)$ becomes important and $E(R) \sim A'' R^{-1}$ ($A''>0$).
In either case $E(R)$ has a minimum, say, 
at $\Rbar$ where $\Rbar$ is the size of the lump.

Of course the radius $R$ cannot be too small in order for the 
nonrelativistic approximation to be valid.  Further the above fermion
lump configuration need to be energetically favored over a configuration
in which fermions reside in the scalar configuration $\phi \sim f_2$.

The above-stated conditions can be satisfied if the mass $m$ of the fermion
in ${\cal E}_f$ differs from the mass $m_0$ in the vacuum  $\phi = f_2$.  
We suppose that $\ep < 0$ inside the lump.  In this situation it is 
energetically favorable for fermions to reside inside the lump rather than
in the boundary region so that $N_F \sim (4\pi R^3/3) \rho_n$.
For a Fermi ball discussed in refs.\ \cite{Campbell, Arafune}
$N_F =N_{f, \rm wall}$ and $\rho_n =0$.
Note that $\ep < 0$ implies that 
\beeq
\rho_n > {\ep_0 \over gf} \sim {\lambda\over g} \, f^3 f_r 
\label{cond1}
\eneq
where $f = \onehalf(|f_1| + f_2)$ and $f_r = (f_2 - |f_1|)/f$.
The nonrelativistic approximation for the fermi gas is justified 
if $mc^2 \rho_n$ dominates over $A \rho_n^{5/3}$ in ${\cal E}_f$;
\beeq
m \gg \rho_n^{1/3}  > 
\Big( {\lambda f_r \over g} \Big)^{1/3} \, f ~.
\label{cond2}
\eneq
We would like to have the anti-de Sitter space inside the lump;
${\cal E}_f + \ep < 0$, or more strongly
${\cal E}_f \ll |\ep |$.  This leads to
\beeq
{|\ep| \over \rho_n}  \sim  gf \gg  m ~~.
\label{cond3}
\eneq

In the examples of the lump solutions  given below we shall
have $|\ep| \gg \ep_0$.  In other words 
$g \rho_n |f_S| \gg V[f_S]$.   Nevertheless $f_S$ is close to $f_1$,
and it is useful to introduce the parameter $h_S = (f_1 - f_S)/f \ll 1$.
$\rho_n$ and $h_S$ are related by $g \rho_n \sim 2 \lambda f^3 h_S$.
The conditions (\ref{cond1}), (\ref{cond2}), and (\ref{cond3}) read
\beeq
2 h_S > f_r \next  
\bigg( {\lambda h_S\over g^4} \bigg)^{1/3} \ll {m\over gf} \ll 1 ~~.
\label{cond4}
\eneq
It also implies that $\rho_n/f^3 \sim 2\lambda h_S/g \ll 2 g^3$.

To satisfy (\ref{cond4}) the fermion
mass inside the lump, $m$, must be much smaller than $gf$.  
If the fermion acquired a mass only from the coupling 
$g\phi \psibar \psi$, then we would have $m \sim gf$.  This apparent
dilemma can be  circumvented if the fermion has Yukawa couplings to more
than one scalar field.  Suppose there are two scalar fields, 
$\phi$ and $\Phi$, which couple to the fermion through
$(g\phi + G\Phi) \psibar \psi$.  In the true vacuum
$\la \phi\ra = f_2$ and $\la \Phi\ra = f_3$.  Hence the fermion 
mass in the true vacuum is $m_0 = g f_2 + G f_3$ whereas 
inside the lump $m = gf_1 + Gf_3$.  We imagine that 
the two terms nearly cancel each other inside the lump such that
$m \ll gf$ and $m_0 \sim 2gf$.  This scenario, at the same time,
provides the stability of fermions inside the lump.  
It costs a huge amount of energy for  a fermion to escape outside the lump.

We stress that although we suppose, to facilitate numerical evaluation, 
that
$m \ll m_0$ in the subsequent discussions, this condition is not
absolutely necessary  for the lump solutions to exist.  In general
cases we need to treat fermion contributions more accurately, taking
into account the back-reaction to the metric as well.

\section{Behavior of solutions}

Having made the assumption described in the previous section, 
we come back to the problem of solving the equations (\ref{metric2}) and 
(\ref{scalar2}).  

\bigskip
\noindent
{\bf (i) Region I}

Near the origin $\phi(r)$ is very close to  $f_S$ which is given by
\beeq
V'[f_S] + \rho_0 = 0 ~~.
\label{fS}
\eneq
$f_S$ itself is close to $f_1$.   Denoting
$\phi(0)$ by $\phi_0$,  we have $0 < \phi_0 - f_S \ll |f_S|$.
The regularity of the solution at the origin leads to 
\beqn
\phi(r) &=&\phi_0+\phi_2 r^2+\cdots ~~, 
 \hskip 1cm \phi_2= \frac{1}{6}
  \bigl( V'[\phi_0]+ \rho_0 \bigr) ~, \cr
\noalign{\kern 5pt}
p(r) &=& 1+p_4r^4+\cdots ~~, \hskip 1.2cm 
  p_4= - 4\pi G \phi_2^2  \cr
\noalign{\kern 5pt}
M(r)&=&  m_3 r^3 + \cdots ~~,  \hskip 1.7cm 
 m_3 = \frac{4}{3} \pi \Big( V[\phi_0] + \rho_0 \phi_0 \Big) \cr
\noalign{\kern 5pt}
H(r)&=&1- 2G m_3 r^2 + \cdots  ~~.
\label{origin1}
\eeqn

In region I the spacetime is approximately anti-de Sitter.   As $\phi \sim f_S$, 
\beeq
T_{00} = \ep =V[f_S]+\rho_0 f_S < 0\next
H = 1 + {r^2\over a_f^2} \next a_f = \sqrt{ {-3\over 8\pi G \ep}} \next
p=1 ~.
\label{metric2a}
\eneq
We have supposed that ${\cal E}_f \ll |\ep|$. 
The equation for $\phi(r)$ can be linearized in
$\delta \phi(r) = \phi(r) - f_S $.  In terms of $z \equiv r^2/a_f^2$,
\beqn
\Bigg\{ z(1+z) {d^2\over dz^2} 
+ \bigg( {3\over 2} + {5\over 2} z \bigg) {d\over dz}
- {1\over 4} \omega^2 a_f^2 \Bigg\} \, \delta\phi = 0 ~,
\label{scalar4}
\eeqn
where $\omega^2 = V''[f_S]$.  This is Gauss' hypergeometric equation.
The solution  regular at $r=0$ is
\beqn
\delta \phi(r) = \delta \phi (0) \cdot
F( \hbox{$\frac{3}{4}$} + \kappa , \hbox{$\frac{3}{4}$} -  \kappa ,
  \hbox{$\frac{3}{2}$} ; -z) ~,
\label{phi1}
\eeqn
where
\beqn
&&\hskip -1.cm
\kappa = \onehalf \sqrt{ \omega^2 a_f^2 + \hbox{${9\over 4}$} }   ~.
\eeqn
We shall soon see that a solution with lump structure appears for
$\omega a_f \gg 1$ with a particular choice of $\delta \phi(0)$.
The ratio of $\delta \phi'(r)$ to $\delta \phi(r)$ is given by
\beeq
{\delta \phi'(r) \over \delta \phi(r)}
= {4r\over 3 a_f^2} \Big( \kappa^2 - {9\over 16} \Big)
{F( \hbox{$\frac{7}{4}$} +  \kappa , \hbox{$\frac{7}{4}$} -  \kappa ,
  \hbox{$\frac{5}{2}$} ; -z)  \over 
F( \hbox{$\frac{3}{4}$} +  \kappa , \hbox{$\frac{3}{4}$} -  \kappa ,
  \hbox{$\frac{3}{2}$} ; -z) } 
\equiv {2r\over a_f^2} ~ J(z)~.
\label{phi2}
\eneq

The deviation from $f_S$ at the origin, $\delta\phi(0)$, must be
very small in order to have  an acceptable solution.  The behavior of the
hypergeometric function for $\kappa \gg 1$ and $z>0$ 
  is given by 
\cite{Bateman}
\beqn
&&\hskip -1.cm
F(a+ \kappa, a - \kappa, c; -z) 
\sim {\Gamma(c)\over 2 \sqrt{\pi} } ~ 
\kappa^{{1\over 2} - c} ~
z^{- {c\over 2} + {1\over 4} } (1+z)^{{c\over 2} - {1\over 4} -a } ~
 \exp \left\{ 2\kappa \sinh^{-1} \sqrt{z} \right\} ~.
\label{geometricF}
\eeqn
The ratio $\delta\phi(r)/\delta \phi(0)$ grows exponentially   
as $(4\kappa)^{-1} z^{-1/2} (1+z)^{-1/4} \, 
\exp \left\{ 2\kappa \sinh^{-1} \sqrt{z} \right\}$.
Near $r=R$, $\delta\phi/|f_S|$ must be very small for the linearization
to be valid.  The ratio  of $F'(z)$ to $F(z)$, $J(z)$ in
(\ref{phi2}),  is given by
\beeq
J(z) \sim {\kappa \over \sqrt{ z(1+z) } } \hskip 1cm
 \hbox{for } \kappa \gg 1 ~,~ z > 0~.
\label{geometricF2}
\eneq

\begin{figure}[tb]
\begin{center}
\includegraphics[height=5cm]{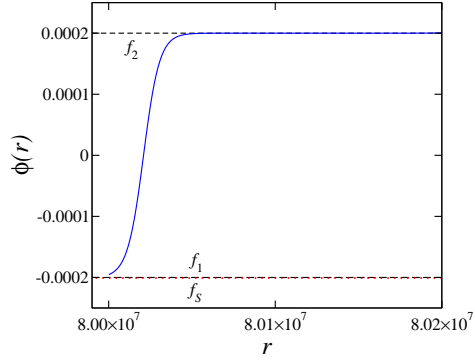}
\end{center}
\caption{$\phi(r)$ of a solution with $f/M_{\Pl}= 0.0002 ,
 f_r= 0.0002 , \lambda = 1$, $h_S=(f_1-f_S)/f=0.005$  and 
$R_1/l_{\Pl}=8 \times 10^7$.  $\phi$
and $r$ are in   units of
$M_{\Pl}$ and $l_{\Pl}$, respectively.}
\label{phi-r}
\end{figure}

\begin{figure}[tb]
\begin{center}
\includegraphics[height=5cm]{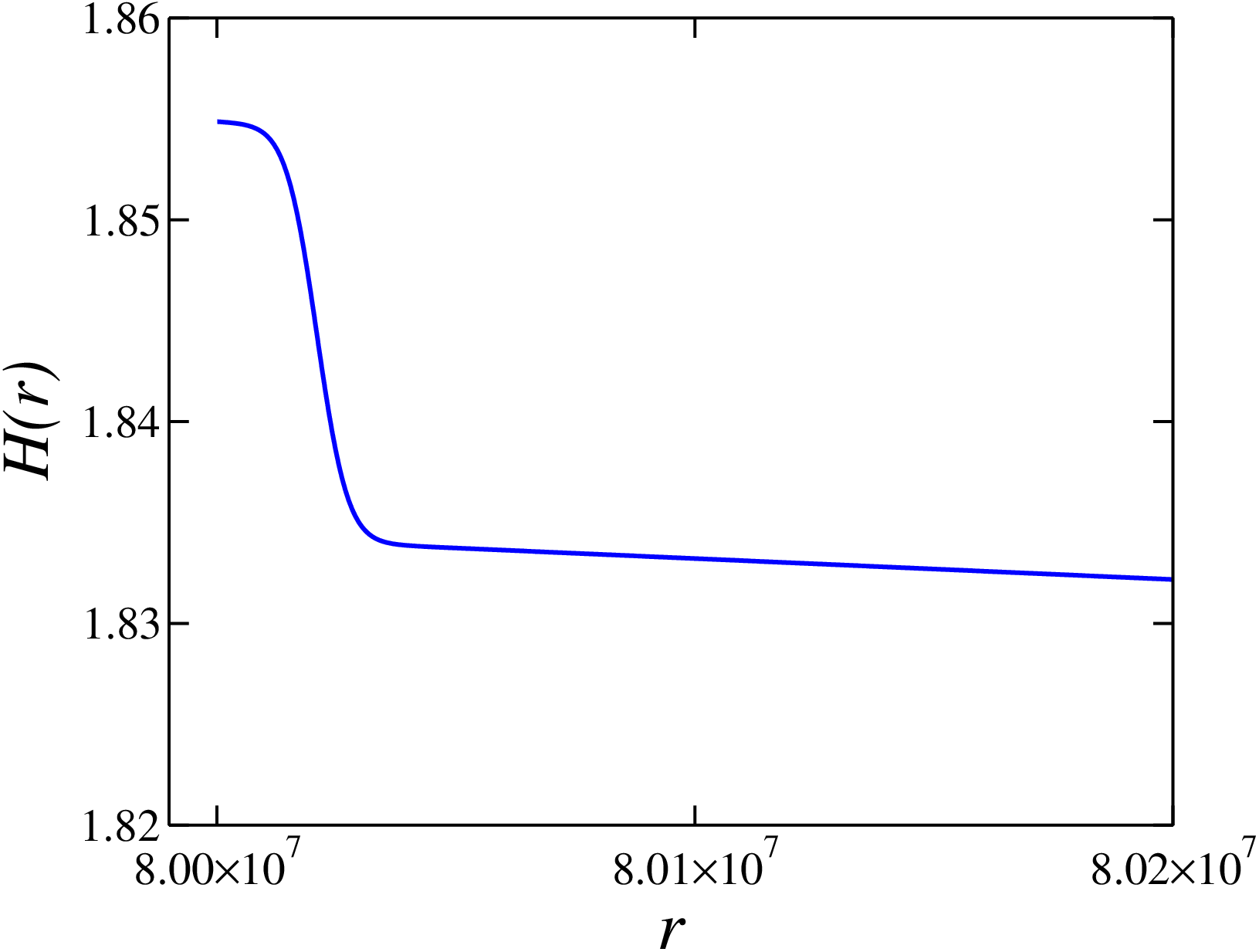}
\end{center}
\caption{$H(r)$ of a solution with $f/M_{\Pl}= 0.0002 ,
 f_r= 0.0002 , \lambda = 1$, $h_S=(f_1-f_S)/f=0.005$ and $R_1/l_{\Pl}=8
\times 10^7$.  $r$ is in   units of $l_{\Pl}$.}
\label{H-r}
\end{figure}

\begin{figure}[tb]
\begin{center}
\includegraphics[height=5cm]{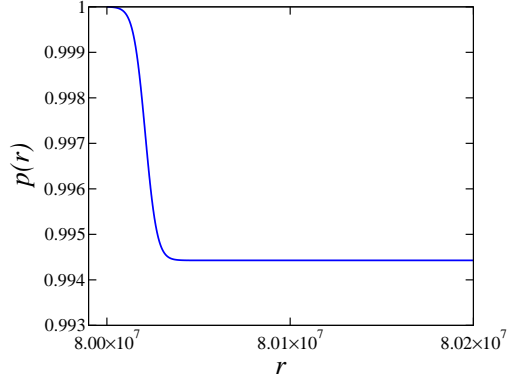}
\end{center}
\caption{$p(r)$ of a solution with $f/M_{\Pl}= 0.0002 ,
 f_r= 0.0002 , \lambda = 1$, $h_S=(f_1-f_S)/f=0.005$ and $R_1/l_{\Pl}=8
\times 10^7$.  $r$ is in   units of $l_{\Pl}$.}
\label{p-r}
\end{figure}

\begin{figure}[tb]
\begin{center}
\includegraphics[height=5cm]{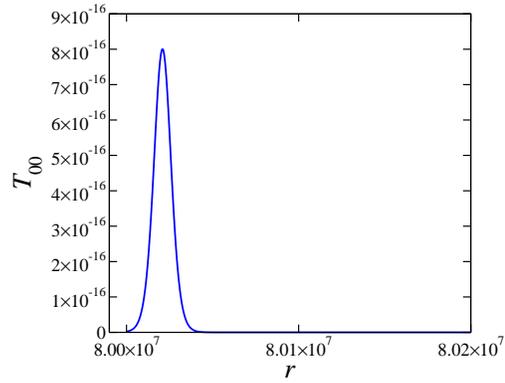}
\end{center}
\caption{The energy density $T_{00}$  for a solution with $f/M_{\Pl}=
0.0002 , f_r= 0.0002 , \lambda = 1$, $h_S=(f_1-f_S)/f=0.005$ and
$R_1/l_{\Pl}=8 \times 10^7$. 
$T_{00}$ and $r$ are in the units of $M_{\Pl}^4$ and $l_{\Pl}$, 
respectively.  For $r<R_1$, $T_{00} \sim \ep < 0$.}
\label{T_00-r}
\end{figure}

\begin{figure}[tb]
\begin{center}
\includegraphics[height=5cm]{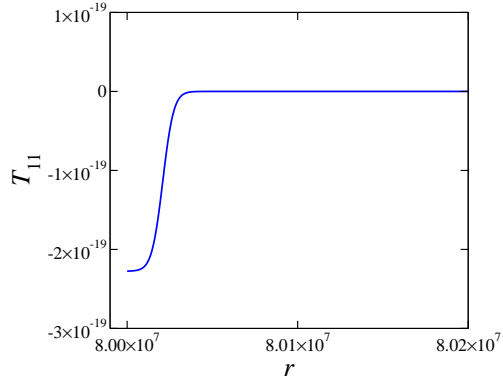}
\end{center}
\caption{The radial pressure $T_{11}$  for a solution with  $f/M_{\Pl}=
0.0002 , f_r= 0.0002 , \lambda = 1$, $h_S=(f_1-f_S)/f=0.005$ and
$R_1/l_{\Pl}=8 \times 10^7$. 
$T_{11}$ and $r$ are in the units of $M_{\Pl}^4$ and $l_{\Pl}$, 
respectively. For $r<R_1$, $T_{11} \sim - \ep > 0$.}
\label{T_11-r}
\end{figure}

\bigskip
\noindent
{\bf (ii) Region II}

In region II, $\phi(r)$ varies substantially and
the nonlinearity of  the equations plays an essential role.
In this region the equations must be solved numerically.  With fine
tuning of the value of $\delta\phi(R_1)$ nontrivial lump solutions
are found.

The algorithm for numerically finding solutions is the following.  First
$R_1$,
$\delta
\phi(R_1)$ and
$f_S$ are chosen and $\delta\phi'(R_1)$ is evaluated 
using (\ref{phi2}) and (\ref{geometricF2}).  In this paper
we  investigate the solutions in the case that $|f_1-f_S| \ll f$ and 
the linearization in Region I is valid.
To a good approximation the metric is given by 
 $H(R_1) = 1 + (R_1/a_f)^2$ and $p(R_1)=1$.  With the boundary conditions
$\delta\phi(R_1)$ and $\delta \phi'(R_1)$  the equations are  numerically
solved in region II. The width of the transition region,
$w = R_2 - R_1$ is approximately given by 
$1/\sqrt{\lambda} \, f$.\cite{Hosotani1}

The  behavior of solutions in region II is displayed in Fig.
\ref{phi-r}.  When the   values of the input parameters are
chosen to be
 $\lambda=1$, $f/M_{\Pl} = 0.0002$, $f_r=\Delta f/f = 0.0002$ and
$h_S=(f_1- f_S)/f =0.005$, then the output parameters
are
$\rho_0 l_\Pl^3 =8.0593\times 10^{-14}$,
 $\epsilon /M_\Pl^4 =-1.5944\times10^{-17}$, $|\ep|/\ep_0 = 74.738$, 
$a_f/l_{\Pl}=8.6525\times 10^7$ and
$\kappa=1.23275 \times 10^4$.  
Here $l_{\Pl}$ is the Planck length.
We note that $\rho_0/f^3 = 0.0101$, $\ep/f^4= - 0.00997$,  and $a_f f =
17305$. For $R_1/l_{\Pl}=8\times 10^7$ ($R_1/a_f = 0.924589$)
we find a solution with 
$\delta\phi(R_1)/M_{\Pl}=5.545749 \cdots \times 10^{-6}$.  
In this example, the value of
$\delta\phi$ at the origin ($r=0$) is found from Eq. (\ref{phi1}) to be
$9.5\times10^{-8857}$, which explains why  numerical integration of
 $\phi(r)$  from $r=0$ to $R_1$ is not possible.
A small discontinuity
in $\phi''$ appears at $r=R_1$ due to the discontinuous change in
$\rho_S(r)$.

The field $\phi(r)$ approaches $f_2$ for $r > R_2$.  In the numerical
integration $R_1$ and $f_S$ are kept fixed while  $\delta\phi(R_1)$ is
varied.   If $\delta\phi(R_1)$ is taken to be slightly smaller, then 
$\phi(r)$ starts  to deviate from $f_2$ in the negative direction, 
approaching $f_1$ as $r$ increases.  If $\delta\phi(R_1)$ is taken to be
slightly larger, then $\phi(r)$ overshoots $f_2$, diverging to $+\infty$ as
$r$ increases. With just the right value of $\delta\phi(R_1)$, the
spacetime becomes nearly flat in region III.

\bigskip
\noindent
{\bf (iii) Region III}

The behavior of the solution in region III is easily inferred.
  From the numerical integration in region II both $H_2=H(R_2)$
and $p_2 = p(R_2)$ are determined.  In region III the metric can be
written in the Schwarzschild form 
\beqn
&&\hskip -1.cm
H(r) = 1 - {2 G \widetilde M \over r} \cr
&&\hskip -1.cm
p(r) = p_2  ~~, 
\label{metric5}
\eeqn
where $\widetilde M$ is the mass of the lump.  
As $\phi(r)$ approaches its asymptotic value $f_2$ and  $T_{00}$
vanishes,  $H(r)$ must take this form.  The value of $\widetilde M$ is
determined numerically either by integrating $T_{00}$ in 
Eq.\ (\ref{metric2}), or 
by fitting $H(r)$ in Eq.\ (\ref{metric2}) just outside
the shell.  In the numerical example presented above,  $\widetilde M<0$.  
As   can be seen in fig.\ \ref{H-r}, $H(r)$   converges to the 
asymptotic value from above.  A rough estimate of $\widetilde M$ is
\beeq
\widetilde M  \sim {4\pi \over 3} R_1^3 ~\ep 
\sim - {M_\Pl^3\over 10 f^2\sqrt{\lambda h_s}} 
\bigg( {R_1\over a_f} \bigg)^3 ~~~.
\label{mass-estimate}
\eneq

\section{Scales of the solutions}

In the previous section we have solved the linearized field equation for
$\phi(r)$ in  region I where $\rho_S \not= 0$ and the deviation from
$f_S$ was found to be small. We numerically determined the behavior in the
nonlinear region II. The resulting structure is a lump with  negative
mass and fermions at its core.  In this section we present more detailed
numerical results.  As discussed in  the previous section the boundary
between regions I and II, located at the matching radius $R_1$, is rather
arbitrary, subject only to the condition that the linearization be
accurate 
 up to that radius.  Precise tuning is necessary for $R_1$ and 
$\delta\phi(R_1)$ in order to obtain a lump solution.  Technically it is
easier to keep $R_1$ fixed and adjust
$\delta\phi$ at $R_1$.   If $\delta\phi(R_1)$ is chosen too small,  $\phi$
comes back toward $f_1$, but cannot reach it,   eventually oscillating
about
$\phi=0$ as $r$  increases.  If $\delta\phi(R_1)$ is chosen too large, 
$\phi$ overshoots $f_2$ and continues to increase.   With just the right
value of $\delta\phi(R_1)$, $\phi$ will approach  $f_2$.  
There can appear shell solutions in which $\phi$ goes back to $f_1$ at 
$r=\infty$.  Such solutions are discussed in ref.\
\cite{Hosotani1}. 

One example of the lump solutions is displayed in Fig.\ \ref{phi-r} for
the parameter values $f/M_{\Pl} =0.0002$, $f_r =0.0002$,  $\lambda=1$ and
$h_S=(f_1-f_S)/f=0.005$.  If we take $R_1/l_\Pl =8\times10^7$ 
($R_1/a_f = 0.924589$),
$\delta\phi(R_1)$ musto be fine-tuned to more than 7 digits:
$\delta\phi(R_1)/M_\Pl \sim 5.545749 \cdots \times10^{-6}$.  
In the transition region both $H(r)$ and $p(r)$ decrease in a one-step
fashion.   See Figs. \ref{H-r} and \ref{p-r}. 
Inside  the lump $H(r)$ is given by (\ref{metric2a}), whereas outside the 
lump it is given by 
\beeq
H = 1 - {2 G \widetilde M \over r}  
\label{metric6}
\eneq
where $\widetilde M /M_\Pl \approx - 3.337\times10^7$
($\widetilde M /f \approx - 1.669 \times10^{11}$). 
$p(r)$ assumes the
constant values 1 and 0.9944 inside and
outside the lump, respectively.

The behavior of the 
energy-momentum tensors, $T_{00}=- T_{22}=-T_{33}$ and $T_{11}$, are 
displayed in Fig.\ \ref{T_00-r} and Fig.\ \ref{T_11-r}.  The energy density 
$T_{00}$ has one sharp peak associated with the rapid variation of $\phi$.  
The radial pressure $T_{11}$  is $\sim -\ep >0$ for $r<R_1$ and becomes
negative at $R_1$. It
increases to zero quickly, and it remains zero outside the lump.  The
absolute value of
$T_{11}(R_1)$ in the transition region II is very small ($\sim 2.3 \times
10^{-19} M_{\Pl}^4$) compared with the maximum value of 
$T_{00}$ ($\sim 8 \times 10^{-16} M_{\Pl}^4$).  
The contributions of the  kinetic energy, $\onehalf H \phi'^2$, and potential
energy, $V[\phi]$, almost  cancel each other in the transition region.

The model contains several dimensionless parameters, $\lambda$, $f/M_{\Pl}$,
$f_r$, $R_1/a_f$ and $\rho_0 f^3$.  When the values of these
parameters are varied, the size of the resulting lump solutions also varies.
In the numerical evaluation we took values of  
$f/M_{\Pl}$ and $f_r$ in the range  between $10^{-4}$ and $10^{-3}$.  It is of
great interest to determine the structure of the lump solutions  when, for
instance, $f/M_{\Pl}= 10^{-19}$.  One can  obtain insight into this problem
by investigating the dependence of the solutions on the above parameters.

\begin{figure}[tb]
\begin{center}
\includegraphics[height=5cm]{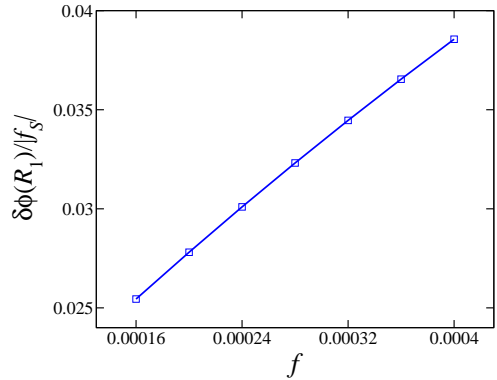}
\end{center}
\caption{The $f$ dependence of $\delta \phi (R_1)$.  $f_r=0.0002$, 
$R_1/a_f=0.9$ and
$h_S=(f_1-f_S)/f=0.005$ are fixed.  $f$ is in  a unit of $M_{\Pl}$.}
\label{f}
\end{figure}

In the numerical investigation  values of $\lambda$, $f/M_{\Pl}$,
$f_r$, $R_1/a_f$ and $h_S$ are given, and we  try to find a  desired value
of $\delta\phi(R_1)$ for a solution to exist.  
In Fig. \ref{f}, $\delta \phi (R_1)/|f_S|$
is plotted as a function of $f$.  It is seen that as $f$ increases, the
value of $\delta \phi$ at $R_1$ needs to be increased in order to get a
lump solution.   The numerical evaluation becomes unreliable 
 when $\delta \phi (R_1)/|f_S|$ becomes large and the
linearization in Eq.\ (\ref{scalar2}) is no longer valid.  

\begin{figure}[tb]
\begin{center}
\includegraphics[height=5cm]{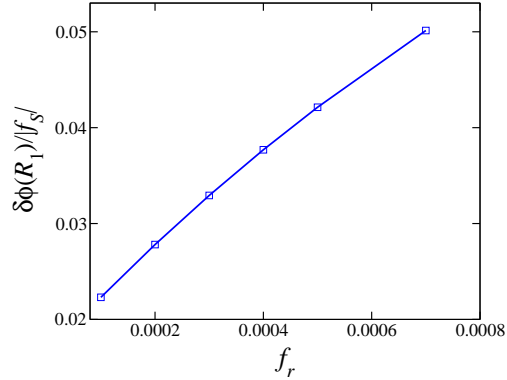}
\end{center}
\caption{The $f_r$ dependence of $\delta \phi (R_1)$.  $f=0.0002$, 
$R_1/a_f=0.9$ and
$h_S=(f_1-f_S)/f=0.005$ are fixed.  $f_r$ is in a  unit of $M_{\Pl}$.}
\label{f_r}
\end{figure}

\begin{figure}[tb]
\begin{center}
\includegraphics[height=5cm]{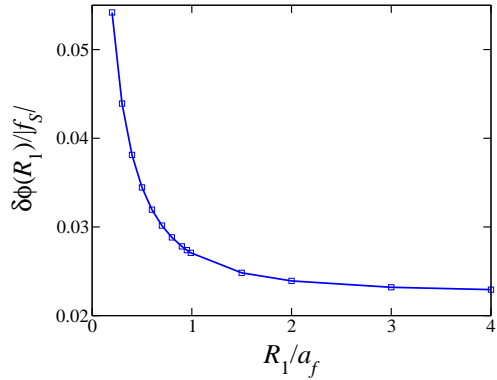}
\end{center}
\caption{The $R_1$ dependence of $\delta \phi (R_1)$.  
$f/M_{\Pl}=0.0002$, $f_r=0.0002$
and $h_S=(f_1-f_S)/f=0.005$ are fixed.}
\label{R_1}
\end{figure}

In Fig. \ref{f_r}, $\delta \phi (R_1)/|f_S|$ is plotted as a function of
$f_r$.  We find that $\delta \phi (R_1)/|f_S|$ increases as $f_r$
increases.  In Fig. \ref{R_1}, $\delta
\phi (R_1)/|f_S|$ is plotted versus $R_1/a_f$.  As $R_1/a_f$ increases, 
$\delta \phi (R_1)/|f_S|$ decreases and approaches a constant
($\sim0.022$).

\begin{figure}[tb]
\begin{center}
\includegraphics[height=5cm]{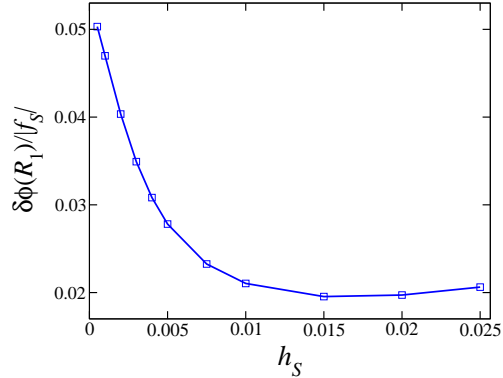}
\end{center}
\caption{The $h_S$ dependence of $\delta \phi (R_1)$.  
$f/M_{\Pl}=0.0002$, $f_r=0.0002$
and $R_1/a_f=0.9$ are fixed.}
\label{f_S}
\end{figure}

\begin{figure}[tb]
\begin{center}
\includegraphics[height=5cm]{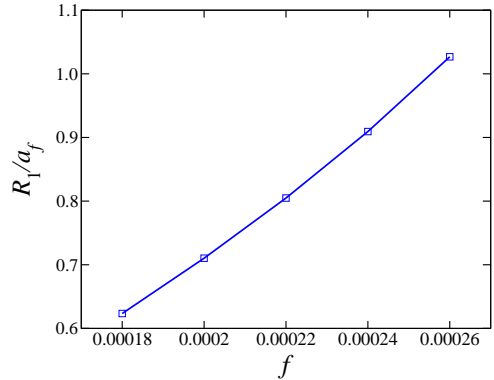}
\end{center}
\caption{$R_1$ versus $f$.  $f_r=0.0002$, $h_S=(f_1-f_S)/f=0.005$ and
$\delta\phi(R_1)/|f_S|=0.03$ are fixed.  
$f$ is in a unit of $M_{\Pl}$.}
\label{size1}
\end{figure}

In Fig. \ref{f_S}, $\delta
\phi (R_1)/|f_S|$ is plotted versus $h_S$.  In this figure $\delta \phi
(R_1)/|f_S|$ decreases as $h_S$ increases and then it begins to increase when
 $h_S$ becomes greater than approximately $0.015$.  The
reason for the increase in the value of $\delta \phi (R_1)/|f_S|$ is that at
some point $|f_S-f_1|$ becomes greater than $\delta\phi(R_1)$, which means
that $\phi(R_1)<f_1$.  In this situation the solution will diverge to
negative infinity if we do not take
$\delta\phi(R_1)$ large enough.  In fig.\ \ref{size1} 
$R_1$  is plotted as a function of $f$ with
 other parameters, including $\delta\phi(R_1)/|f_S|$, kept fixed. 

\begin{figure}[tb]
\begin{center}
\includegraphics[height=5cm]{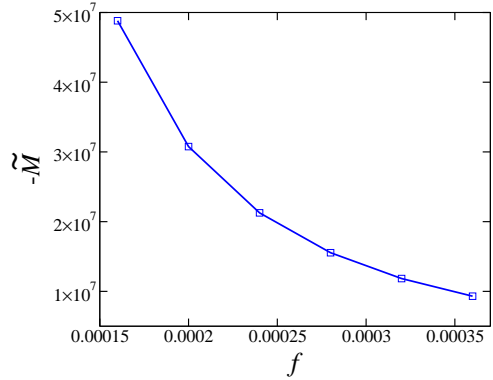}
\end{center}
\caption{$\widetilde M$ versus $f$.  $f_r=0.0002$, $h_S=(f_1-f_S)/f=0.005$
and
$R_1/a_f=0.9$ are fixed.  $f$ and  $\widetilde M$ are in a unit of 
$M_{\Pl}$.}
\label{size2}
\end{figure}

\begin{figure}[tb]
\begin{center}
\includegraphics[height=5cm]{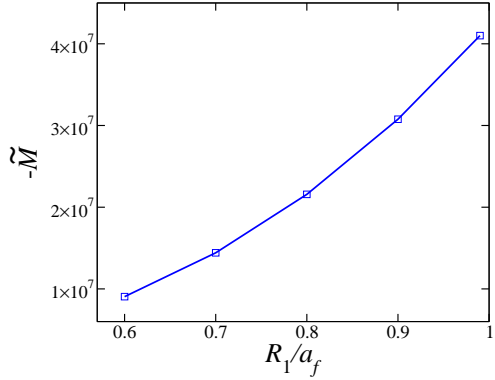}
\end{center}
\caption{
$\widetilde M$ versus $R_1$.  $f/M_{\Pl}=0.0002$, $f_r=0.0002$ and 
$h_S=0.005$   are fixed.  $\widetilde M$ is in a unit of $M_{\Pl}$.}
\label{size3}
\end{figure}

With  the above results  we can obtain good information concerning
 the mass  $\widetilde M$ of the lumps.  In Figs.\ \ref{size2} and 
 \ref{size3}  $\widetilde M$ is plotted as a function of $f$ and $R_1$,
respectively.  We find that the estimate (\ref{mass-estimate}) is
good. As an example, let us take $f_r=0.0002$, $h_S=0.005$,
$\lambda=0.01$, and $g=1$.  Then $a_f \sim 1.2 \cdot 10^5 \cdot 
(1 \, {\rm GeV}/f)^2 \, {\rm m}$ and
$- \widetilde M \sim 4.6 \cdot 10^{31} \cdot (1 \, {\rm GeV}/f)^2  \,
(R_1/a_f)^3 \, {\rm kg}$.  As $f$ becomes small,
the size of the lump becomes very large.

\section{Discussion}

In this paper we have demonstrated that there 
exist fermionic gravitating lumps with a false vacuum core
of the scalar field.  This is a curious structure which may appear at
various  scales.  We have constructed   lump solutions for the case 
$h_S = (f_1 - f_S)/f  > f_r = (f_2- |f_1|)/f$, 
which implies that $\lambda f_r/g  < \rho_n/ 2f^3 \ll g^3$.  
 (See  Eq.\ (\ref{cond4}).)
If the typical energy scale is very high, the size
of the lump will be small but it can be produced abundantly in the 
early universe.  If the energy scale becomes low, 
this size may increases to a cosmic scale.  

We have focused on the cases in which fermion contributions
or back-reaction to the metric is small.  Certainly this 
restriction needs to be relaxed in order to accommodate more realistic
models in the particle theory.  The lump solutions presented in this
paper have negative masses as seen outside the lumps.  As the mass and
density of fermions inside the lump increase, the mass of the lump will
become positive. The solution probably continues to exist in this case,
though explicit construction  of such a solution is necessary by taking
account of the back-reaction to the energy-momentum tensors.   An
investigation along this line is  in progress, and we hope to 
 report it in the  near future.

A false vacuum state may appear in the standard model of 
electroweak and gravitational interactions.  In ref.\ 17) it is
 shown that the false vacuum of the Higgs field emerges in the early
universe as a result of the  winding of the gauge fields in the 
$S^3 \times R^1$ Robertson-Walker spacetime.  As quarks and leptons have
relevant Yukawa couplings to the Higgs field, gravitating  lumps
with quarks and leptons at the core  may be copiously produced.
As the universe expands, the barrier separating the false vacuum from the
true vacuum disappears so that the gravitating fermionic lumps would 
become unstable, the fermions inside the lumps dissipating to  
infinity.  It is of great interest to elucidate the consequences of this
process.

Furthermore, in the higher-dimensional gauge theory defined on orbifolds,
false vacua naturally appear in the gauge field configurations.\cite{HHHK} 
It would be interesting  to investigate if the gauge interactions of
fermions produce gravitating lumps, as we have found for the Yukawa
interactions.

\vskip 1.cm

\section*{Acknowledgments}

This work was supported in part  by Scientific Grants 
from the Ministry of Education and Science of Japan, Nos.\ 13135215 
 and   13640284.  The very early stage of this investigation 
 was carried out with the help of Masafumi Hashimoto
whose contribution is gratefully acknowledged.

%%%%%%%%%%%%%%%%%%%%%%%%%%%%%%

% A useful Journal macro
\def\jnl#1#2#3#4{{#1}{\bf #2} (#4), #3}

\def\Zphys{{\em Z.\ Phys.} }
\def\jssc{{\em J.\ Solid State Chem.\ }}
\def\jpsJ{{\em J.\ Phys.\ Soc.\ Japan }}
\def\ptps{{\em Prog.\ Theoret.\ Phys.\ Suppl.\ }}
\def\PTP{{\em Prog.\ Theoret.\ Phys.\  }}

\def\JMP{{\em J. Math.\ Phys.} }
\def\NPB{{\em Nucl.\ Phys.} B}
\def\NP{{\em Nucl.\ Phys.} }
\def\PLB{{\em Phys.\ Lett.} B}
\def\PL{{\em Phys.\ Lett.} }
\def\PRL{\em Phys.\ Rev.\ Lett. }
\def\PRB{{\em Phys.\ Rev.} B}
\def\PRD{{\em Phys.\ Rev.} D}
\def\PR{{\em Phys.\ Rev.} }
\def\PRe{{\em Phys.\ Rep.} }
\def\AP{{\em Ann.\ Phys.\ (N.Y.)} }
\def\RMP{{\
em Rev.\ Mod.\ Phys.} }
\def\ZPC{{\em Z.\ Phys.} C}
\def\SCI{\em Science}
\def\CMP{\em Comm.\ Math.\ Phys. }
\def\MPLA{{\em Mod.\ Phys.\ Lett.} A}
\def\IJMPB{{\em Int.\ J.\ Mod.\ Phys.} B}
\def\cmp{{\em Com.\ Math.\ Phys.}}
\def\JPA{{\em J.\  Phys.} A}
\def\CQG{\em Class.\ Quant.\ Grav. }
\def\ATMP{{\em Adv.\ Theoret.\ Math.\ Phys.} }
\def\ibid{{\em ibid.} }

%%%%%%%%%%%%%%%%%%

\end{document}